%% file: paper.tex
\documentclass[twocolumn,showpacs,aps,prl,superscriptaddress,floatfix,nofootinbib]{revtex4}
\usepackage{graphicx}
\usepackage{verbatim}
\usepackage{dcolumn}
\usepackage{amsmath}
\usepackage{epsfig}
\usepackage{subfigure}
\usepackage{soul}
\usepackage{xcolor}

\newcommand{\BaBarYear}      {21}
\newcommand{\BaBarNumber}    {002}
\newcommand{\BaBarType}      {PUB}  
\newcommand{\SLACPubNumber}  {17608}

\input babarsym.tex
\def\MeV{\mev}
\def\GeV{\gev}

\begin{document}

\pagestyle{plain}

\begin{flushleft}
\babar-\BaBarType-\BaBarYear/\BaBarNumber \\
SLAC-PUB-\SLACPubNumber\\
\end{flushleft}

\title{{\large \bf Search for Darkonium in $e^+e^-$ Collisions}}

\input authors_mar2021_frozen

\begin{abstract}
Collider searches for dark sectors, new particles interacting only feebly with ordinary matter, have largely focused 
on identifying signatures of new mediators, leaving much of dark sector structures unexplored. In particular, the existence 
of dark matter bound states (darkonia) remains to be investigated. This possibility could arise in a simple model in 
which a dark photon ($A'$) is light enough to generate an attractive force between dark fermions. We report herein a 
search for a $J^{PC}= 1^{--}$ darkonium state, the $\Upsilon_D$, produced in the reaction 
$e^+e^- \rightarrow \gamma \Upsilon_D, \Upsilon_D \rightarrow A' A' A'$, where the dark photons subsequently decay into pairs 
of leptons or pions, using $514\text{ fb}^{-1}$ of data collected with the {\babar\ }detector. No significant signal is observed, and we set bounds on the $\gamma - A'$ kinetic mixing as a function of the dark sector coupling constant for $0.001 < m_{A'} < 3.16 \GeV$ and $0.05 < m_{\Upsilon_D} < 9.5 \GeV$. 
\end{abstract}

\pacs{12.60.-i,14.80.-j, 95.35.+d}

\maketitle

\setcounter{footnote}{0}
The possibility of dark sectors, new quantum fields neutral under all standard model (SM) forces, has emerged as an 
intriguing framework to explain the presence of dark matter in the universe~\cite{Pospelov:2007mp,ArkaniHamed:2008qn}. 
While these particles don't couple directly to ordinary matter, indirect interactions through low-dimensional operators 
called ``portals'' are possible~\cite{Beacham:2019nyx}. The physics of these dark sectors could involve an arbitrary 
number of fields and interactions, including the possibility of self-interacting dark matter. This scenario can be realized 
in a minimal dark sector model containing a single Dirac fermion ($\chi$) charged under a new $\mathrm{U}(1)$ gauge group 
with a coupling constant $g_D$~\cite{An:2015pva}. The corresponding force carrier is conventionally referred to as a dark 
photon ($A'$), and couples to the SM photon via kinetic mixing with strength $\varepsilon$~\cite{Fayet:1980rr,Holdom:1985ag}. 
A light dark photon would give rise to an attractive force between the $\chi$ and $\bar\chi$ particles, resulting in 
the formation of bound states (darkonia) when $1.68m_{A'} < \alpha_D m_{\chi}$ for $\alpha_D=g_{D}^2/4\pi$~\cite{An:2015pva,Rogers}. 

The two lowest energy bound states in this model are denoted $\eta_D$ ($J^{PC}= 0^{-+}$) and $\Upsilon_D$ ($J^{PC}= 1^{--}$), 
in analogy with similar SM states. The quantum numbers predict the following production and decay mechanisms at $e^+e^-$ colliders: 
$e^+e^- \rightarrow \eta_D + A', \eta_D \rightarrow A' A'$ and initial-state radiation (ISR) process 
$e^+e^- \rightarrow \Upsilon_D + \gamma_{\text{ISR}}, \Upsilon_D \rightarrow A' A' A'$. In the regime $m_{A'} < 2 m_\chi$, 
the dark photon decays visibly into a pair of SM fermions with a decay width proportional to the product $m_{A'}\varepsilon^2$. 
Depending on the value of these parameters, the decays can be prompt or significantly displaced from the $\epem$ interaction 
point. Current constraints on visible $A'$ decays~\cite{Riordan:1987aw,Bjorken:1988as,
Bross:1989mp,Davier:1989wz,Blumlein:2013cua,Curtin:2013fra,Lees:2014xha,Batley:2015lha,Anastasi:2016ktq, 
Banerjee:2018vgk,Aaij:2019bvg} exclude values of $\varepsilon \gtrsim 10^{-3}$ over a wide range of masses from the dielectron 
threshold up to tens of$\,\GeV$~\cite{naturalUnits}.

We report herein a search for darkonium in the ISR process 
$e^+e^- \rightarrow \gamma_{\text{ISR}} \Upsilon_D, \Upsilon_D \rightarrow A' A' A'$, where the dark photons subsequently decay 
into pairs of electrons, muons, or pions. The cross section  is determined for prompt $A'$ decays in the region 
$0.001 \GeV < m_{A'} < 3.16 \GeV$ and $0.05 \GeV < m_{\Upsilon_D} < 9.5 \GeV$. For $m_{A'} < 0.2 \GeV$, the dark photon decay 
length becomes significant for values of $\varepsilon$ we expect to probe, and we additionally report cross sections for 
lifetimes $\tau_{A'}$ corresponding to $c\tau_{A'}$ values of 0.1, 1, and 10 mm. This search is based on 514\invfb of data 
collected with the \babar\ detector at the SLAC PEP-II $e^+e^-$ collider at the $\Y4S$, $\Y3S$, and $\Y2S$ resonances and 
their vicinities~\cite{Lees:2013rw}. The \babar\ detector is described in detail elsewhere~\cite{Bib:Babar,TheBABAR:2013jta}. 
To avoid experimental bias, the data are not examined until the selection procedure is finalized. The analysis is developed using simulated signal events and a small fraction of real data for background studies.

Signal events are generated using MadGraph5~\cite{Alwall:2014hca} with prompt dark photon decays for 119 different 
$A'$ and $\Upsilon_D$ mass hypotheses. For $m_{A'} < 0.2 \GeV$, we also simulate samples with non-zero dark photon lifetimes 
corresponding to proper decay lengths 0.1 mm, 1 mm, and 10 mm. The detector acceptance and reconstruction efficiencies 
are estimated with a simulation based on \textsc{GEANT4}~\cite{Agostinelli:2002hh}. Since the background is too 
complex to be accurately simulated, we use 5\% of the data to optimize the selection criteria, assuming that any 
signal component has a negligible impact on this procedure. This data set, referred to as the optimization sample, 
is discarded from the final results.

The event selection for prompt $A'$ decays proceeds by selecting events containing exactly six charged tracks, and reconstructing 
dark photon candidates as pairs of oppositely charged tracks identified as electrons, muons, or pions by particle identification 
algorithms. We require the presence of at least one lepton pair of opposite charge with the same flavor to limit the large accidental 
background. We form $\Upsilon_D$ candidates by combining three dark photon candidates, and fit them, constraining all tracks to originate 
from a common point compatible with the beam interaction region. For each $\Upsilon_D$ candidate, we additionally form same-sign track 
combinations by swapping particles with identical flavor between reconstructed $A'$ pairs, such as $(e^+e^+)(e^-e^-)(\mu^+\mu^-)$ or 
$(\pi^+\pi^+)(\pi^-\pi^-)(e^+e^-)$. For the fully mixed final state $(\mu^+\mu^-)(\pi^+\pi^-)(e^+e^-)$, we use the same-sign combination $(\mu^+\pi^+)(\mu^-\pi^-)(e^+e^-)$, since pions are more easily misidentified as muons than electrons. The mass difference between 
same-sign pairs, distributed more broadly for signal events than for combinatorial background, is used to further improve the signal 
selection as discussed below.

The detection of the ISR photon accompanying $\Upsilon_D$ production is not explicitly required. Instead, we infer the kinematics 
of the particle recoiling against the $\Upsilon_D$ candidates, and we select the ISR photon candidate that is most compatible with 
the photon hypothesis as follows. If the recoil particle is determined to have been emitted inside the electromagnetic calorimeter acceptance, we search 
for the presence of a corresponding ISR photon candidate, which is defined as a neutral cluster having an energy within 10\% of 
that of the recoiling particle, and an angle compatible with the direction of the recoiling particle to better than 0.1 rad.

To improve the signal purity, we train three multivariate classifiers consisting of logistic regressions stacked on top of Random Forest (RF) classifiers~\cite{Breiman}. The following 13 variables are used as inputs to the RF: the $\chi^2$ of the constrained fit to the $\Upsilon_D$ 
candidate; combined particle identification information of the six tracks; the maximum mass difference between any pair of $A'$ candidates; the polar 
angle and the invariant mass of the particle recoiling against the reconstructed $\Upsilon_D$ candidate; a categorical feature indicating 
whether the recoiling particle is emitted inside the calorimeter acceptance and if a corresponding ISR photon candidate is found; the sum 
of neutral energy deposited in the electromagnetic calorimeter, excluding the ISR photon candidate; the average of the three dark photon helicity angles~\cite{helicity}; the average of the angles between pairs of dark photons in the $\Upsilon_D$ rest frame; the average of the 
dihedral angles between pairs of dark photons; the average of the three helicity angles of the tracks produced in the $A'$ decays; 
the average of the dark photon decay lengths, defined as the distances between the primary interaction point and the $A'$ decay vertices; 
and the maximum mass difference between same-sign pairs. 

\begin{figure}[htbp]
\centering
\includegraphics[width=0.48\textwidth]{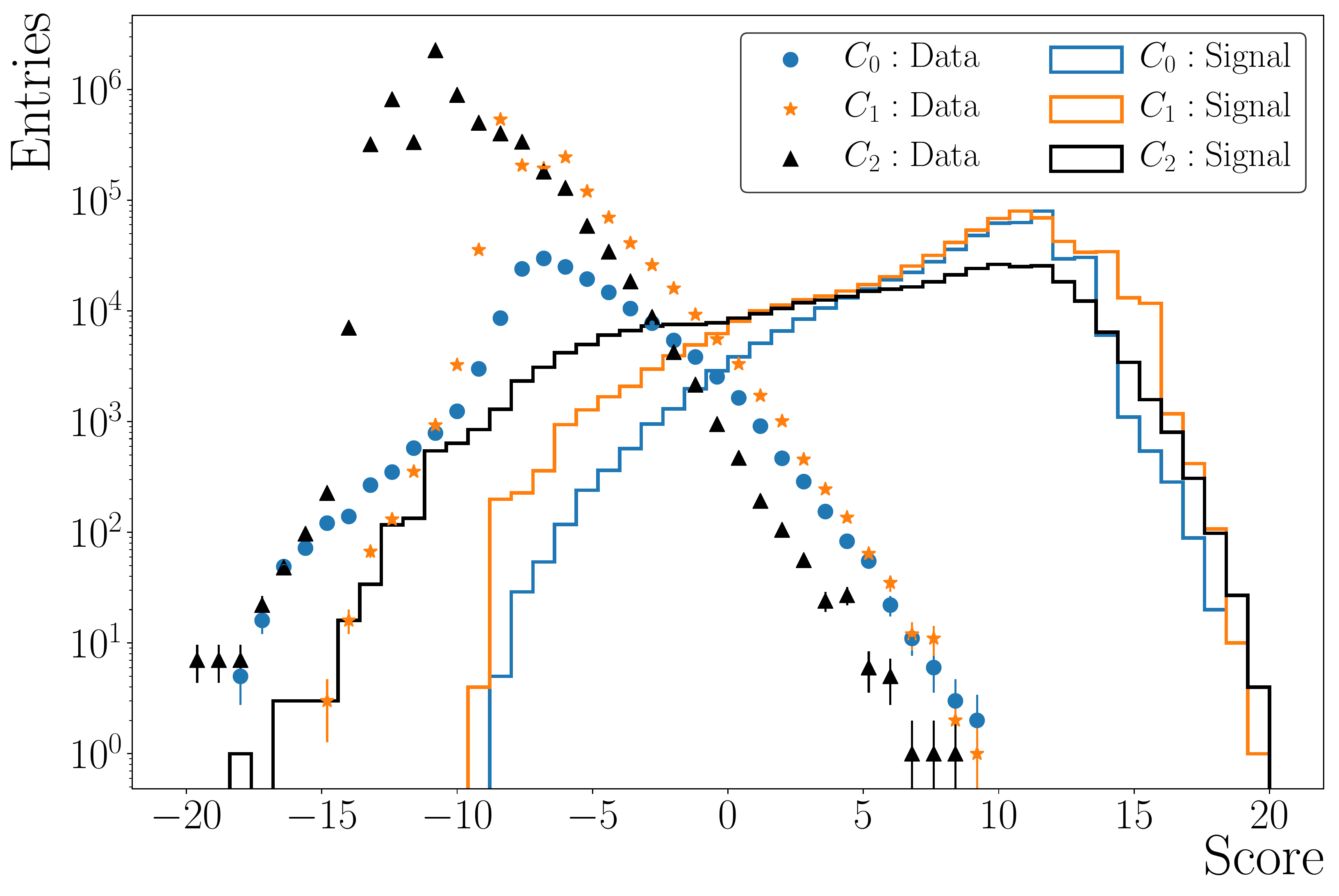}
\caption{The distribution of the classifier scores for each event category for the data (markers) and signal Monte 
Carlo (solid lines) samples. The MC simulations are arbitrarily normalized.}
\label{fig:mlscore_prompt}
\end{figure}

To improve the robustness of the predictions of the classifiers, we group the final states into three categories based on the number 
of pion pairs: zero ($C_{0}$), one ($C_{1}$), or two ($C_2$) pion pairs. A classifier is trained for each category with a sample of 
simulated events for different $\Upsilon_D$ and $A'$ masses and a fraction of the optimization sample to describe the background. The 
classifier outputs are then transformed into classifier scores using a logit function~\cite{logit}, with higher scores indicating greater 
probabilities of being signal events. The distribution of the classifier scores for each category are shown in Fig.~\ref{fig:mlscore_prompt}. 
The optimal selection criteria are determined by maximizing a figure of merit averaged over a wide range of $\Upsilon_D$ and $A'$ 
masses. We adopt a conservative approach and treat observed events as signal candidates for the purposes of calculating the figure 
of merit. If multiple $\Upsilon_D$ candidates are selected in an event, a single one is chosen based on its final state according to 
the following sequence of hypotheses: 6e, 4e2$\mu$, 2e4$\mu$, 6$\mu$, 4e2$\pi$, 2e2$\mu$2$\pi$, 4$\mu$2$\pi$, 2e4$\pi$, 2$\mu$4$\pi$. 

\begin{figure}[htbp]
\centering
\includegraphics[width=0.48\textwidth]{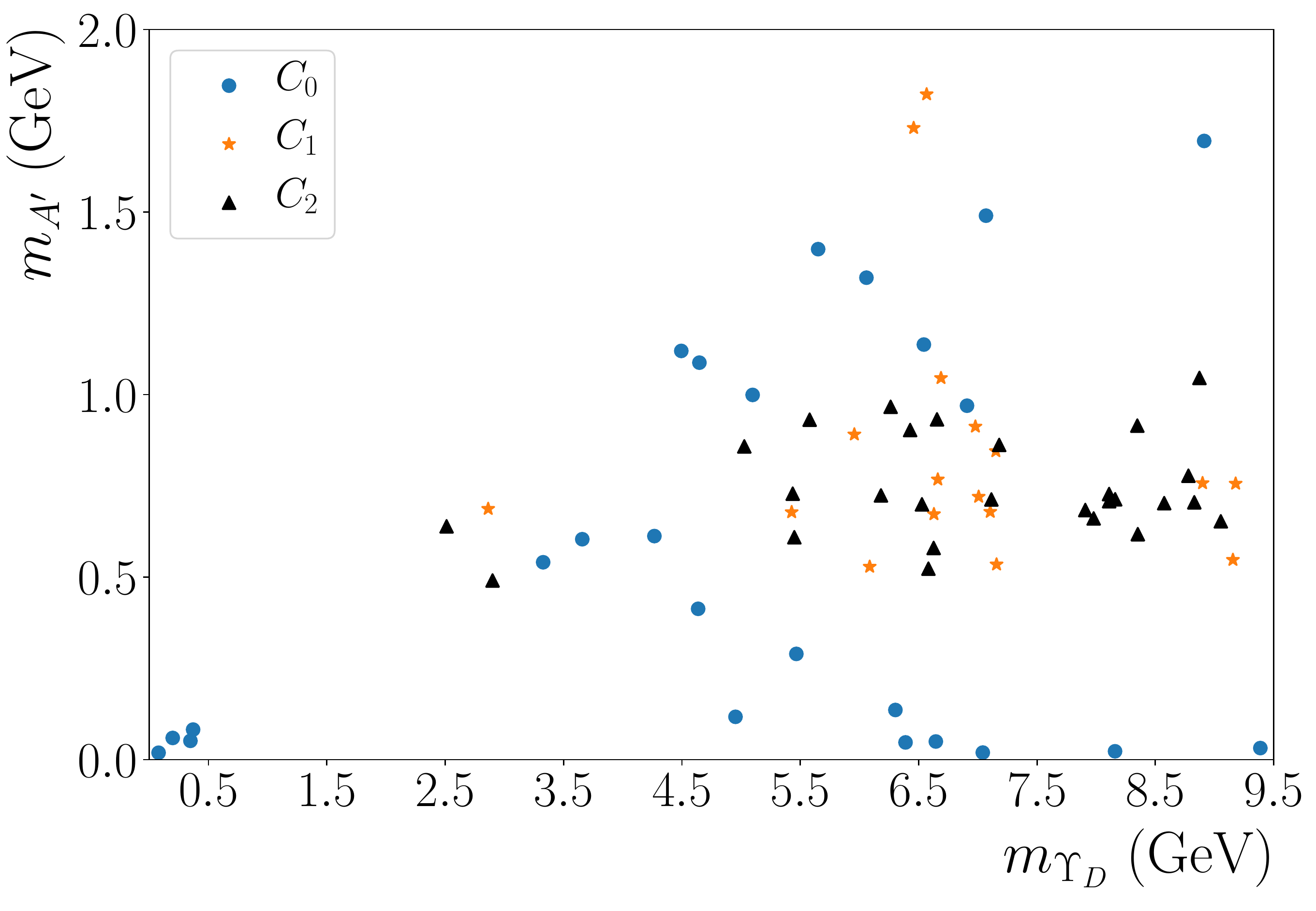}
\caption{The $(m_{\Upsilon_D}, m_{A'})$ distribution for events passing all selection criteria for prompt dark photon decays.}
\label{fig:data_prompt}
\end{figure}

A total of 69 events pass all the selection criteria. The corresponding $(m_{\Upsilon_D}, m_{A'})$ distribution is shown in 
Fig.~\ref{fig:data_prompt}. The events near $m_{\Upsilon_D} \sim 0.1\GeV$ and $m_{A'} \sim 0.05\GeV$ arise from $e^{+}e^{-}\to \gamma \gamma \gamma$ 
events in which all three photons convert to $e^{+}e^{-}$ pairs.

The signal is extracted by combining all event categories into a single sample, and scanning the $(m_{\Upsilon_D}, m_{A'})$ plane 
in steps of the signal resolution. The signal region for a given mass hypothesis is defined as the interval 
$[m_{\Upsilon_D} - 4\sigma_{m_{\Upsilon_D}}; m_{\Upsilon_D} + 4\sigma_{m_{\Upsilon_D}}]$ and 
$[m_{A'} - 4\sigma_{m_{A'}} ; m_{A'} + 4\sigma_{m_{A'}}]$, where $\sigma_{m_{\Upsilon_D}}$ ($\sigma_{m_{A'}}$) denotes the corresponding $\Upsilon_D$ ($A'$) mass resolution. The resolutions are determined by fitting the different signal 
Monte Carlo (MC) samples with a Crystal Ball function~\cite{CrystalBall} and interpolating the results throughout the full mass range. 
The $\Upsilon_D$ ($A'$) mass resolution varies between $5-40\MeV$ ($1-8 \MeV$); the detailed results are available in the 
Supplemental Material~\cite{SPM}. The number of observed background events is estimated by averaging two neighboring regions 
along the $m_{\Upsilon_D}$ axis: $[m_{\Upsilon_D} - 8\sigma_{m_{\Upsilon_D}}; m_{\Upsilon_D} - 4\sigma_{m_{\Upsilon_D}}]$ and 
$[m_{\Upsilon_D} + 4\sigma_{m_{\Upsilon_D}}; m_{\Upsilon_D} + 8\sigma_{m_{\Upsilon_D}}]$. This choice is motivated by the potential 
background contribution due to hadronic resonances or photon conversions, which would be concentrated at similar values of dark 
photon masses. The signal significance is assessed from MC samples, using sideband data from the classifier score 
distribution to model the $(m_{\Upsilon_D}, m_{A'})$ distribution of the background. The most significant measurement contains 
two events in the signal window, corresponding to a p-value of 30\%, which is compatible with the null hypothesis.

\begin{figure}[htbp]
\centering
\includegraphics[width=0.48\textwidth]{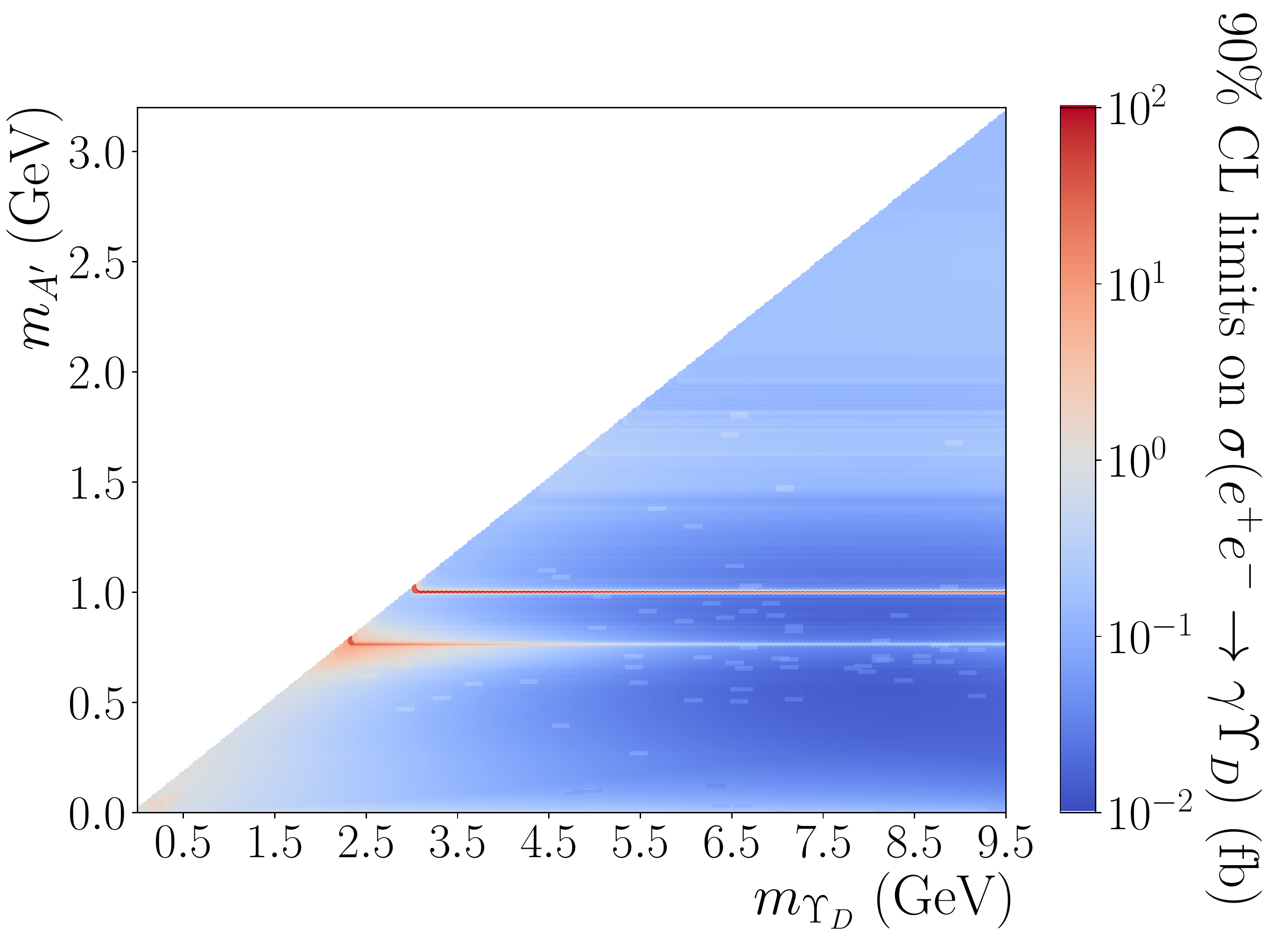}
\caption{The 90\% CL upper limits on the $e^+e^- \rightarrow \gamma \Upsilon_D$ cross section for prompt dark photon decays.}
\label{fig:cross_prompt}
\end{figure}

In the absence of signal, we derive 90\% confidence level (CL) upper limits on the $e^+e^- \rightarrow \gamma \Upsilon_D$ 
cross section using a profile likelihood method~\cite{Rolke:2004mj}. The probability of observing $N$ events in a given signal 
region is described by the following model:
$$P(N | n + b) = \frac{e^{-n} n^{N}}{N!} \frac{e^{-b} b^{B}}{B!} \frac{1}{2\pi\sigma_Z \sigma_L} e^{-\frac{(z-Z)^2}{2\sigma_Z^2}} e^{-\frac{(l-L)^2}{2\sigma_L^2}}$$
where $b$ ($B$) is the expected (estimated) number of background events, and $n = l z \sigma(e^+e^- \rightarrow \gamma \Upsilon_D)$ is 
the expected number of signal events given by the product of the integrated luminosity $l$, the $e^+e^- \rightarrow \gamma \Upsilon_D $ 
cross section, and the signal efficiency $z$. The measured luminosity, signal efficiency, and their uncertainties are denoted by $L$, $Z$, 
$\sigma_L$, and $\sigma_Z$, respectively. The signal efficiency includes the dark photon branching fractions, taken from Ref.~\cite{Batell:2009yf}. 
The efficiency is determined for each simulated sample and interpolated to the full parameter space, ranging from 0.1\% for 
$m_{\Upsilon_D} \sim 0.15\GeV,  m_{A'} \sim 0.05\GeV$ to 34\% for $m_{\Upsilon_D} \sim 8.0\GeV, \, m_{A'} \sim 1.0\GeV$. The uncertainty in 
the signal efficiency arises mainly from particle identification algorithms, assessed with high-purity samples of leptons and pions. This 
source of uncertainty varies between 9\% and 11\%. The uncertainty associated with the efficiency extrapolation procedure ranges from 4\% to 
7\%, depending on the $m_{\Upsilon_D}$ and $m_{A'}$. Other uncertainties include the tracking efficiency (1.2\%) and the limited statistical 
precision of the simulated sample (1\%-5\%). The uncertainty in the dark photon branching fraction~\cite{Batell:2009yf} ranges from parts 
per mille to 1\%. The uncertainty in the luminosity is determined to be 0.6\%~\cite{Lees:2013rw}. The cross section at 90\% CL upper limits are displayed in 
Fig.~\ref{fig:cross_prompt}. The dark photon decays predominantly into $\pi^+\pi^-\pi^0$ ($K^+K^-$) near the $\omega$ ($\phi$) resonance which are not considered in this analysis, 
resulting in much looser bounds around $m_{A'} \sim 0.8 \GeV$ ($m_{A'} \sim  1 \GeV$).
 
\begin{figure}[htbp]
\centering
\includegraphics[width=0.48\textwidth]{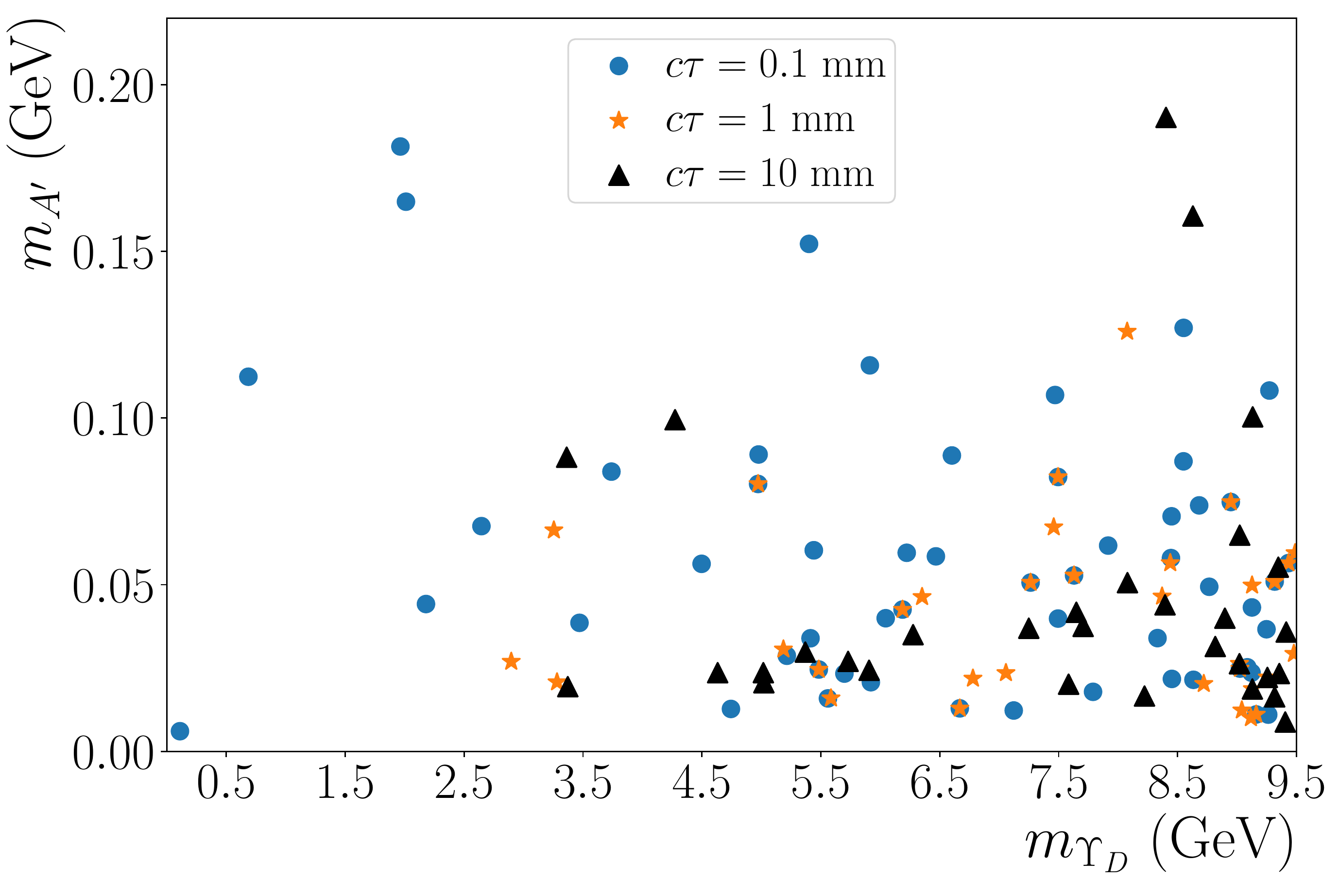}
\caption{The $(m_{\Upsilon_D}, m_{A'})$ mass distribution of event candidates passing all selection criteria for the datasets optimized 
for each dark photon lifetime.}
\label{fig:data_disp}
\end{figure}

We follow a similar procedure to determine the $e^+e^- \rightarrow \gamma \Upsilon_D$ cross section for each dark photon lifetime 
hypothesis. The measurement is performed for $m_{A'}< 0.2 \GeV$. In this mass range, the $A'$ decays almost exclusively to an $\epem$ 
pair. The event selection is analogous to that previously described, except that we constrain the momentum vector of the $A'$ candidates 
to point back to the beam interaction region instead of requiring the tracks to originate from this location when performing the 
$\Upsilon_D$ kinematic fit. To further suppress photon conversions in the detector material, we add the following variables to the RF 
classifier, averaged over the three dark photon candidates: the $\chi^2$ of a fit of the $A'$ candidate; the angle between the secondary 
vertex flight direction and the $A'$ momentum; and the ratio between the flight length and its uncertainty. We train a classifier for 
each $c\tau_{A'}$ value to improve the search sensitivity. A total of 56, 33, and 31 events are selected for the $c\tau_{A'}$ = 0.1, 1, 
and 10 mm data sample, respectively. The resulting mass distributions are shown in Fig.~\ref{fig:data_disp}. The signal extraction procedure 
described above is applied to each selected sample separately. No significant signal is observed for any $A'$ lifetime hypothesis, and 
limits on the cross section for each value of $c\tau_{A'}$ are extracted. The classifier score distributions and the cross section at 90\% CL upper limits are shown in the Supplemental Material~\cite{SPM}. 

\begin{figure}[htbp]
\centering
\includegraphics[width=0.48\textwidth]{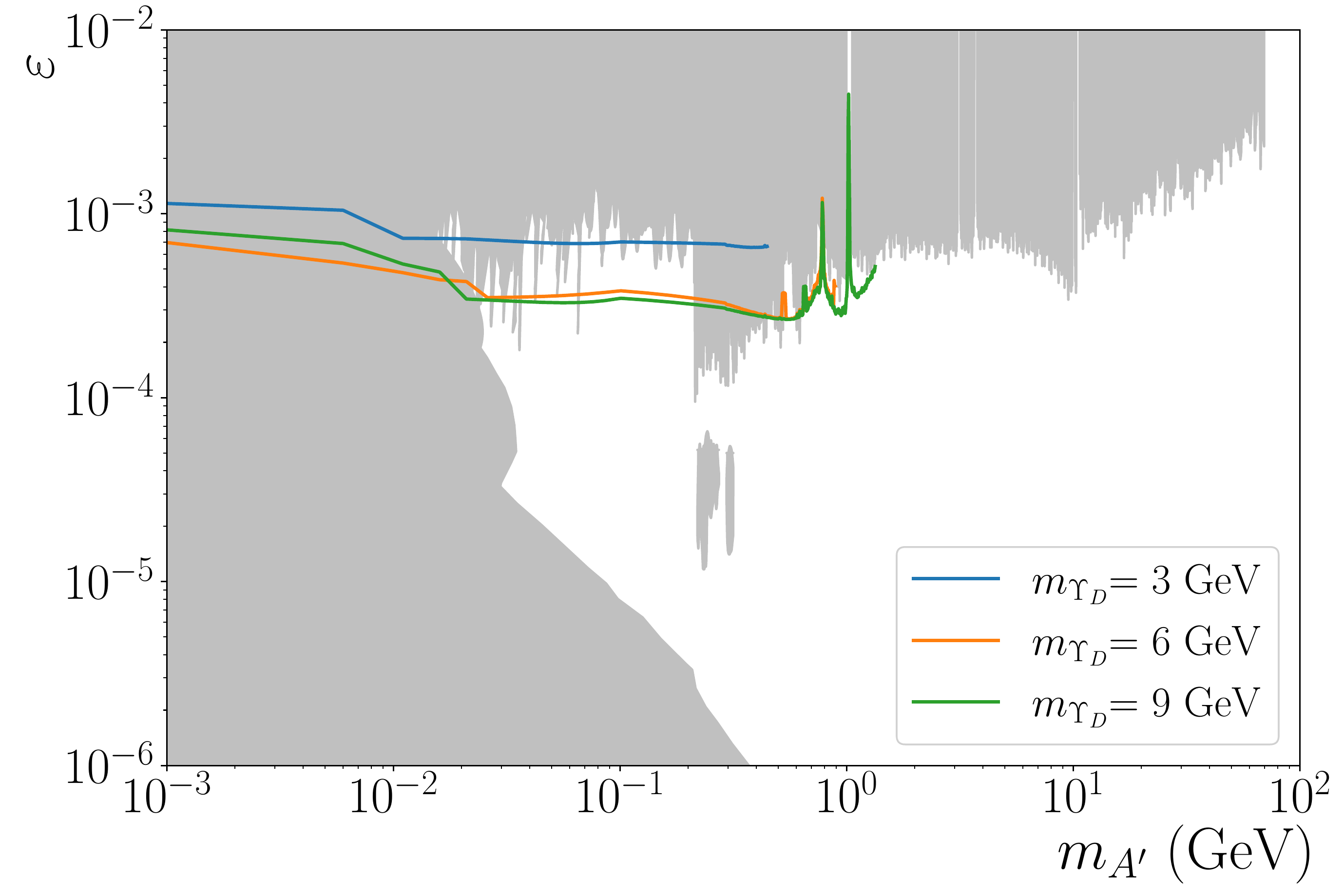}
\includegraphics[width=0.48\textwidth]{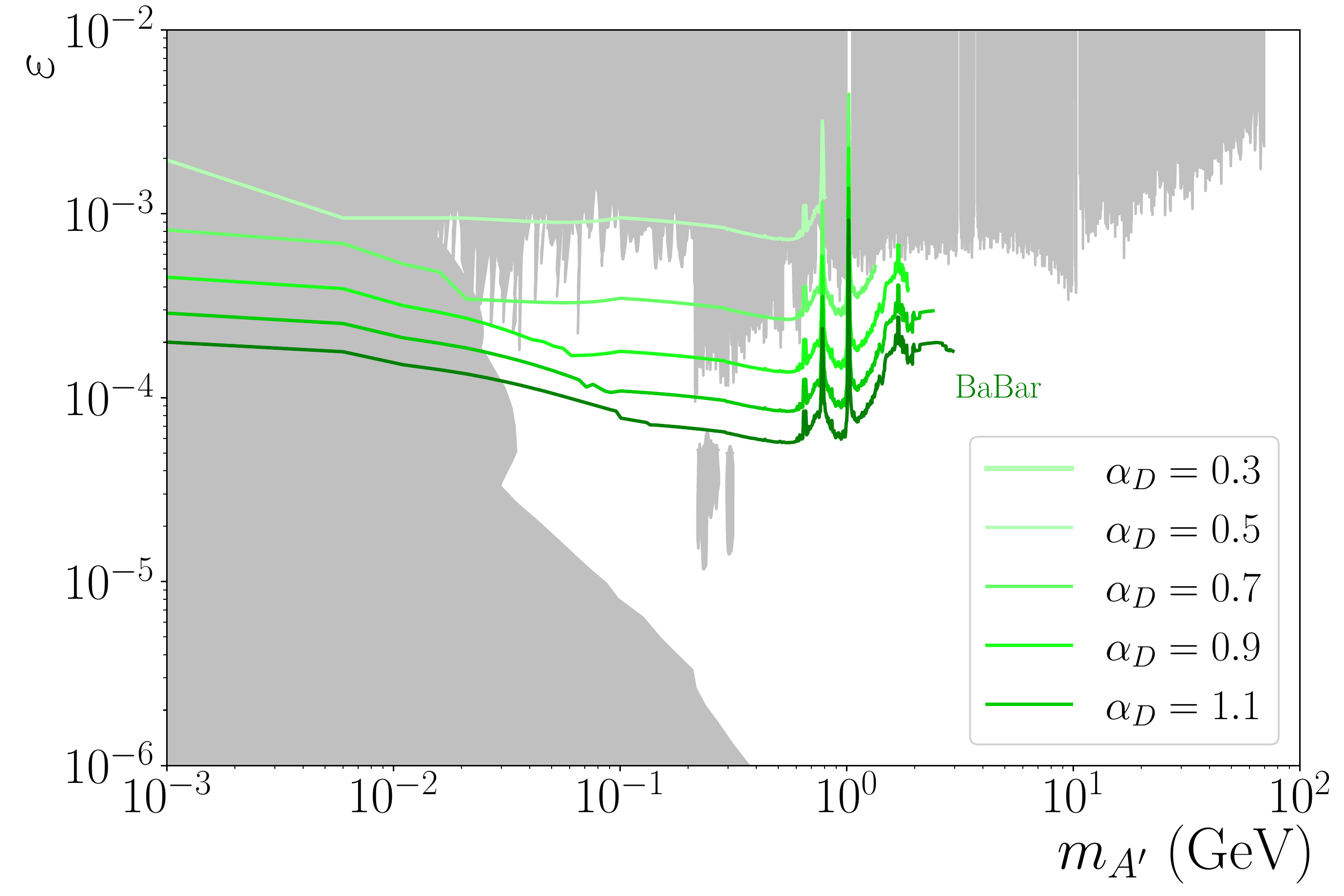}
\caption{The 90\% CL upper limits on the kinetic mixing $\varepsilon$ for (top) various $\Upsilon_D$ masses assuming $\alpha_D=0.5$ 
and (bottom) various $\alpha_D$ values assuming $m_{\Upsilon_D} = 9\GeV$ together with current constraints (gray area)~\protect{~\cite{Riordan:1987aw,Bjorken:1988as,
Bross:1989mp,Davier:1989wz,Blumlein:2013cua,Curtin:2013fra,Lees:2014xha,Batley:2015lha,Anastasi:2016ktq, 
Banerjee:2018vgk,Aaij:2019bvg}}.}
\label{fig:results2}
\end{figure}

The 90\% CL upper limits on the kinetic mixing parameter are extracted by an iterative procedure taking into account the effect of 
the potentially long dark photon lifetime. At each step, we estimate the dark photon lifetime given the current value of the kinetic mixing, 
compare the limit on the production cross section interpolated at that lifetime, update the kinetic mixing, and repeat the procedure until 
convergence is obtained. Since the dark photon lifetime is independent of the dark sector coupling constant, we derive separate limits 
for $\alpha_D$ values set to 0.1, 0.3, 0.5, 0.7, 0.9, and 1.1. The results are shown in Fig.~\ref{fig:results} for $\alpha_D=0.5$, and 
in the Supplemental Material~\cite{SPM} for the remaining values. Bounds on the mixing strength $\varepsilon$ down to $5\times10^{-5} - 10^{-3}$ are 
set for a large fraction of the parameter space. Constraints for different values of $\alpha_D$, $m_{A'}$ and $m_{\Upsilon_D}$ are also 
shown in Fig.~\ref{fig:results2}. 

In summary, we report the first search for a dark sector bound state decaying into three dark photons in the 
range $0.001 \GeV < m_{A'} < 3.16 \GeV$ and $0.05 \GeV < m_{\Upsilon_D} < 9.5 \GeV$. No significant signal is seen, 
and we derive limits on the $\gamma - A'$ kinetic mixing $\varepsilon$ at the level of $5\times10^{-5} - 10^{-3}$, depending on 
the values of the model parameters. These measurements improve upon existing constraints over a significant 
fraction of dark photon masses below $1 \GeV$ for large values of the dark sector coupling constant.

\begin{figure}[htbp]
\centering
\includegraphics[width=0.48\textwidth]{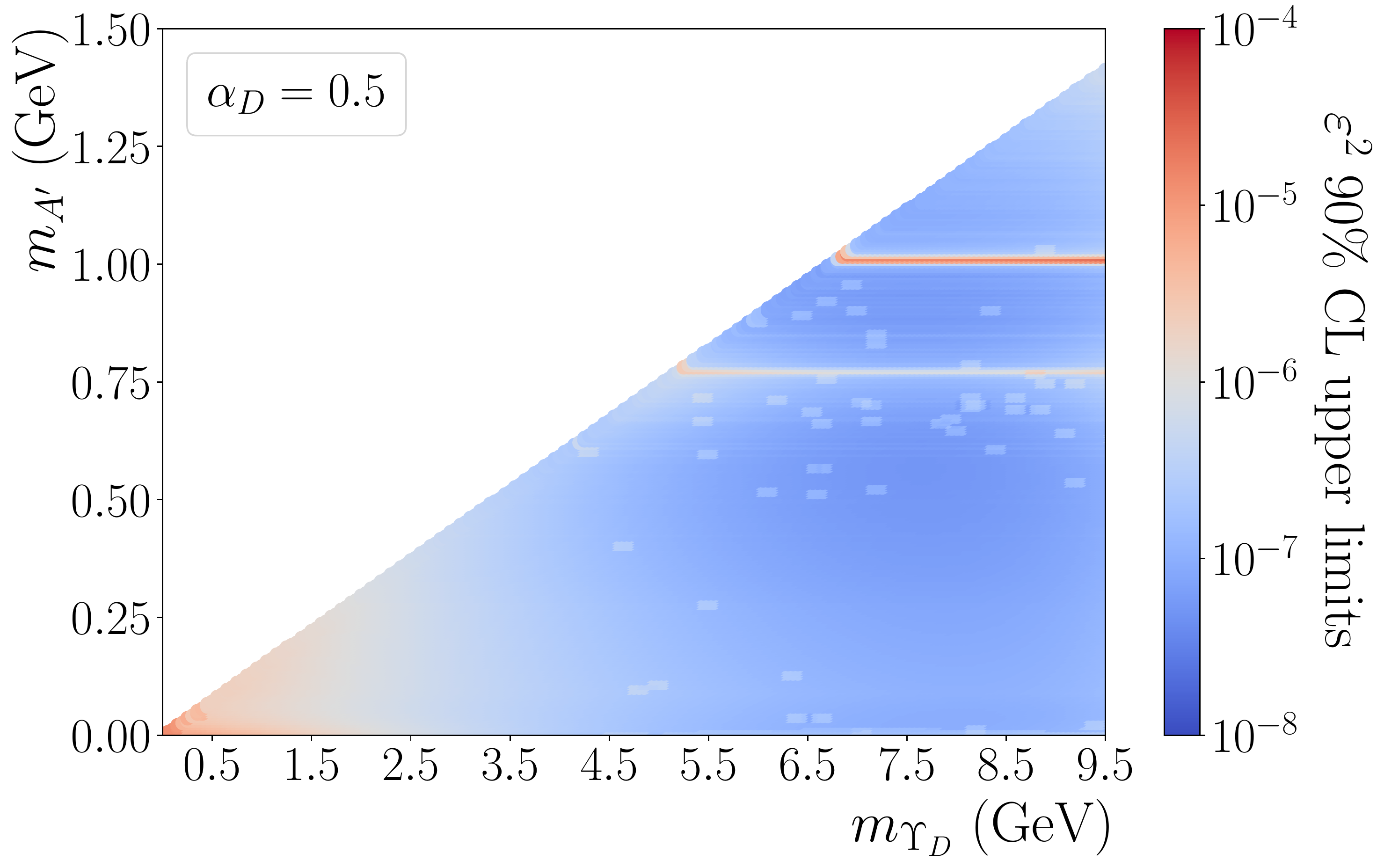}
\caption{The 90\% CL upper limits on the kinetic mixing $\varepsilon^2$ as a function of the $\Upsilon_D$ mass, $m_{\Upsilon_D}$, and 
dark photon mass, $m_{A'}$, assuming $\alpha_D = 0.5$.}
\label{fig:results}
\end{figure}

The authors wish to thank Haipeng An and Yue Zhang for useful discussions and for providing us with MadGraph simulations of self-interacting 
dark matter processes. We also thank Gaia Lanfranchi for providing us constraints from existing experiments.
\input acknowledgements.tex

\clearpage
\onecolumngrid
\begin{center} {\bf \large Supplemental Material for \babar-\BaBarType-\BaBarYear/\BaBarNumber} \end{center}
\begin{center} {\it \large Search for Darkonium in $e^+e^-$ Collisions} \end{center}
\vspace{0.5 cm}

Additional figures for the dark sector bound state search are presented in this Supplemental Material.

\begin{figure}[htbp]
\centering
\includegraphics[width=0.48\textwidth]{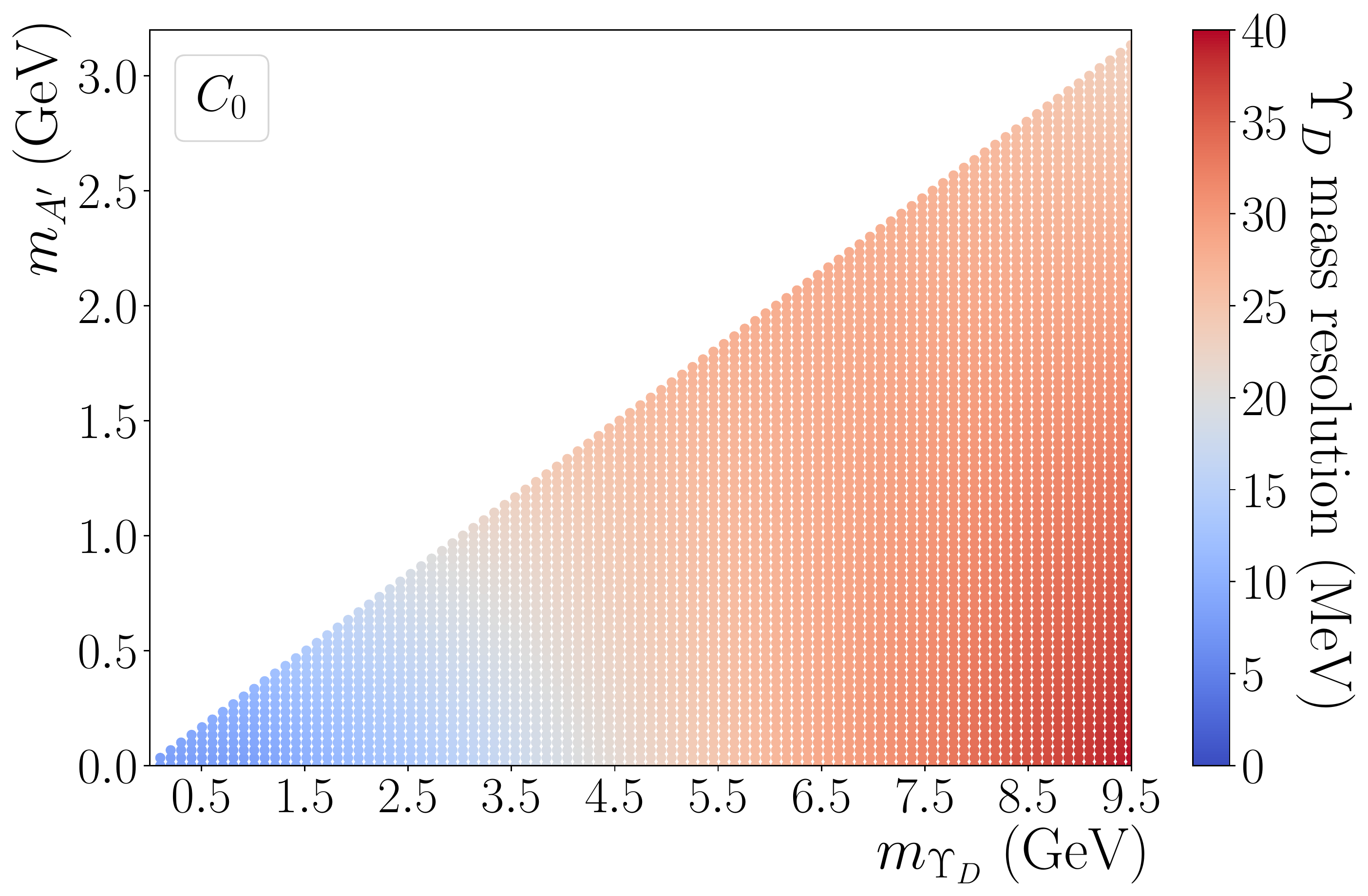}
\includegraphics[width=0.48\textwidth]{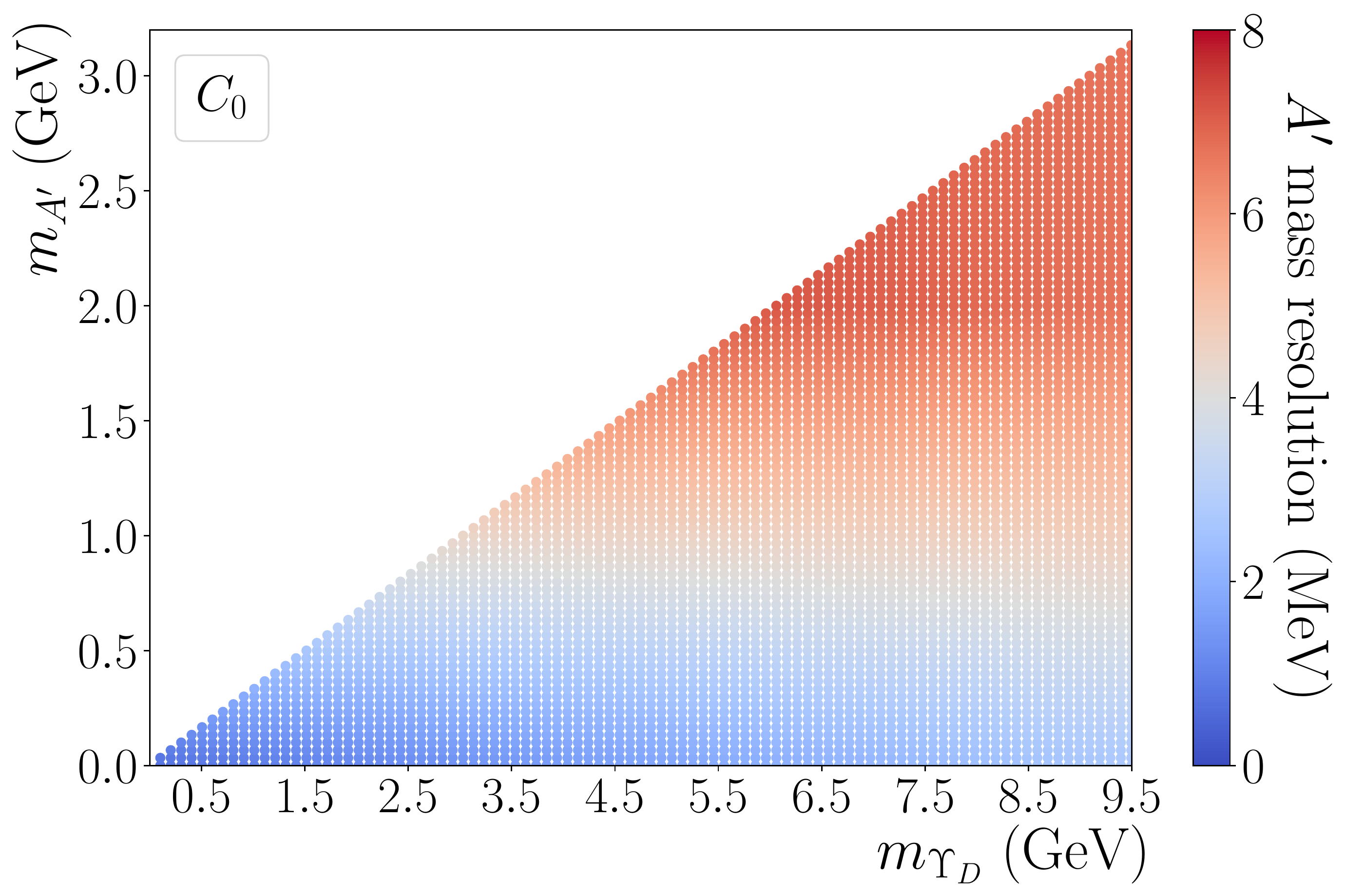} \\ 
\includegraphics[width=0.48\textwidth]{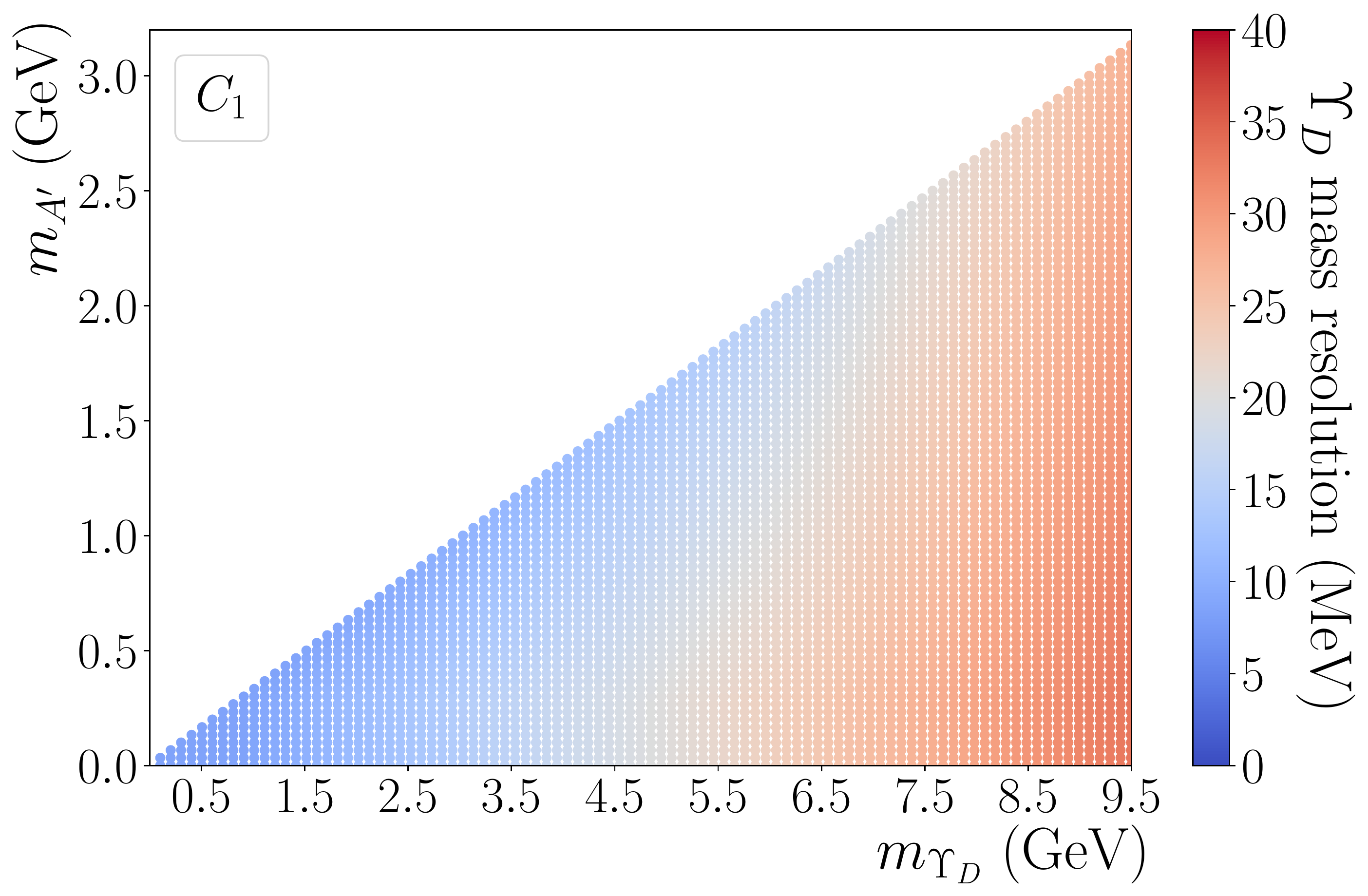}
\includegraphics[width=0.48\textwidth]{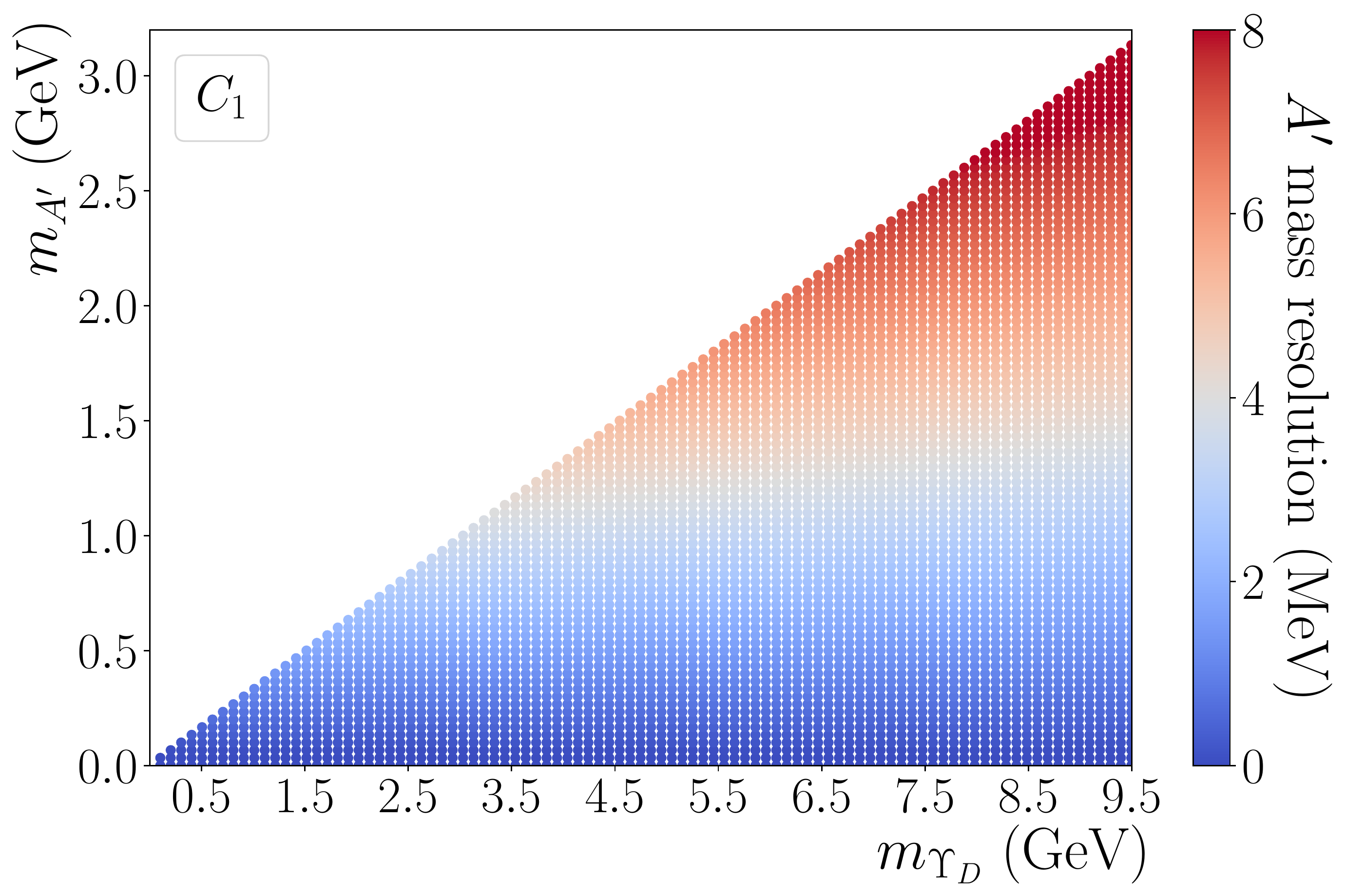} \\ 
\includegraphics[width=0.48\textwidth]{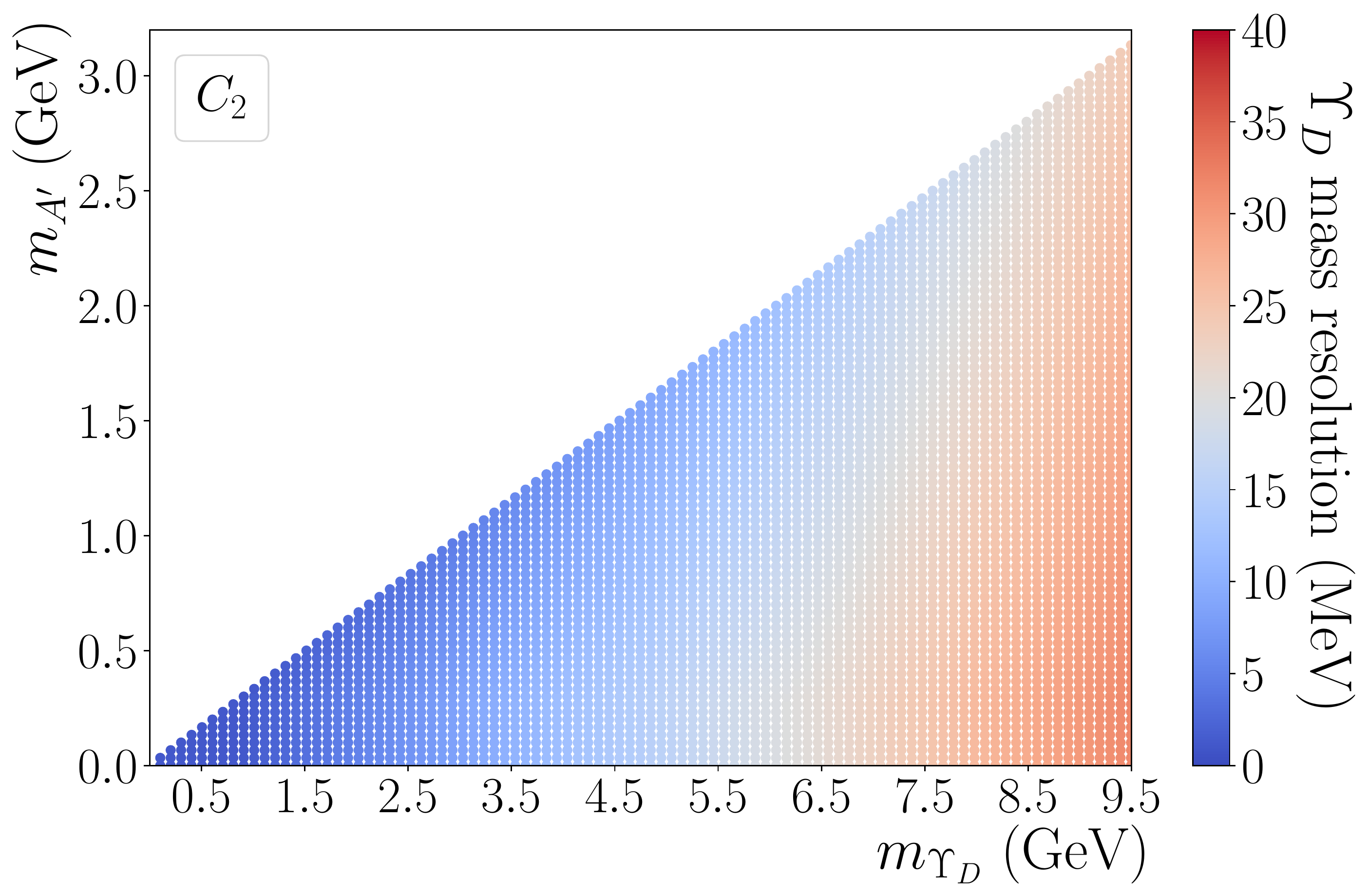}
\includegraphics[width=0.48\textwidth]{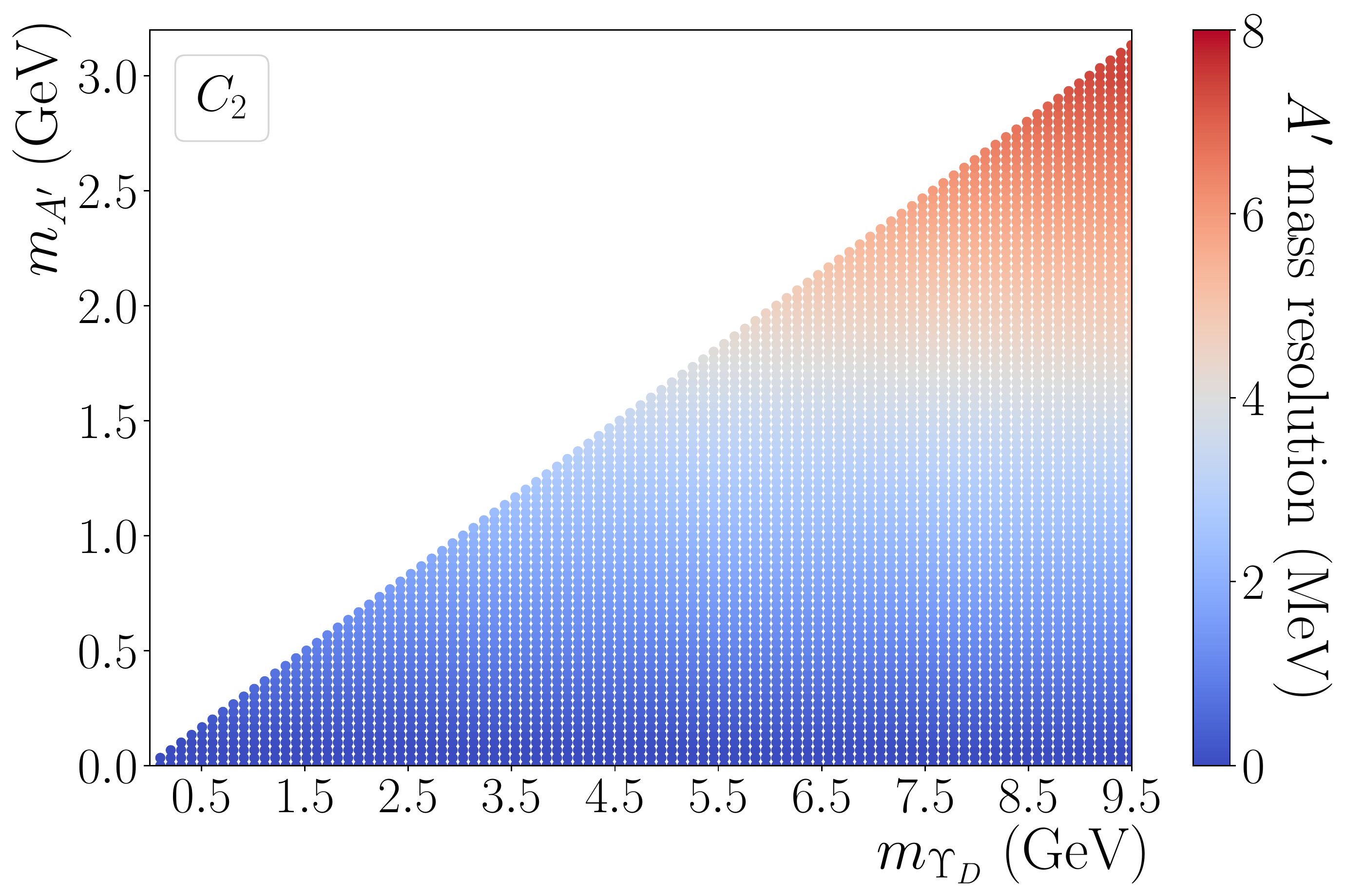} \\ 
\caption{The $\Upsilon_D$ and $A'$ mass resolution for each category of events (see text) for prompt $A'$ decays.}
\label{fig:mass_resolution}
\end{figure}

\begin{figure}[htbp]
\centering
\includegraphics[width=0.48\textwidth]{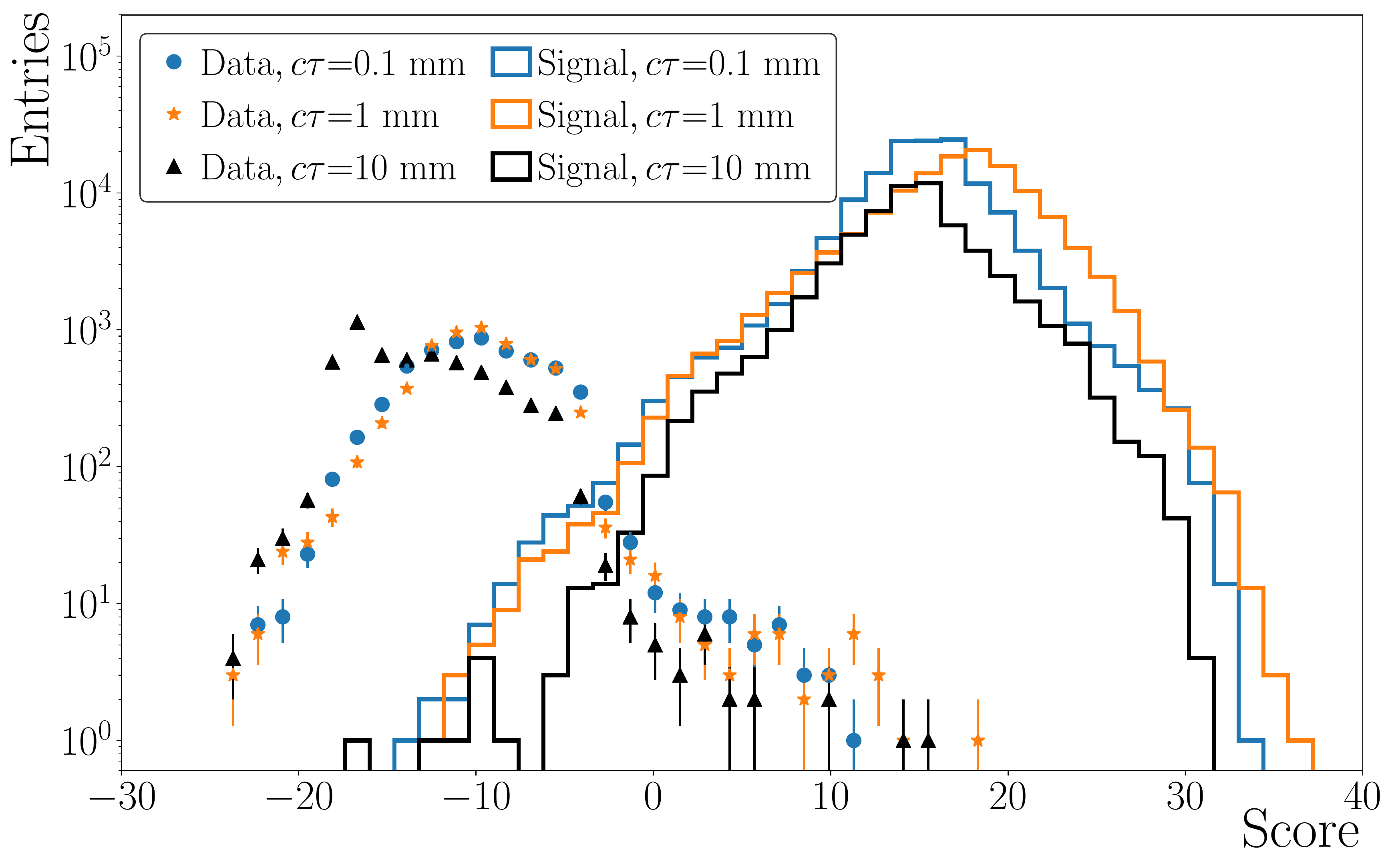}
\caption{The distribution of the classifier scores for data (markers) and signal MC simulations (soild lines) for dark photon lifetimes corresponding to (top) $c\tau_{A'}=0.1 \rm \, mm$, (middle) $c\tau_{A'}=1 \rm \, mm$, and (bottom) $c\tau_{A'}=10 \rm \, mm$. The MC simulations are arbitrarily normalized.}
\label{mlscore_prompt}
\end{figure}

\begin{figure}[htbp]
\centering
\includegraphics[width=0.48\textwidth]{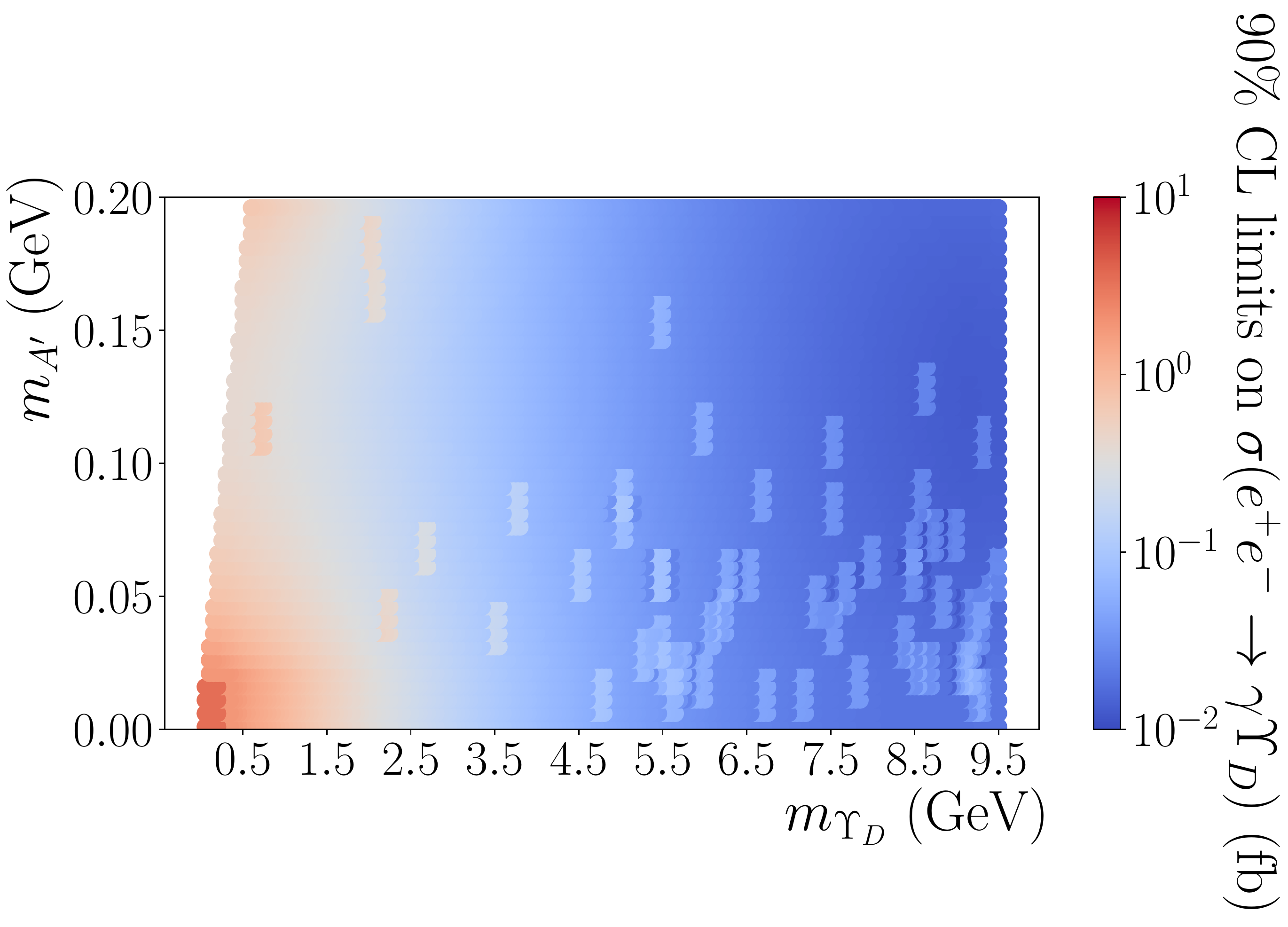}
\includegraphics[width=0.48\textwidth]{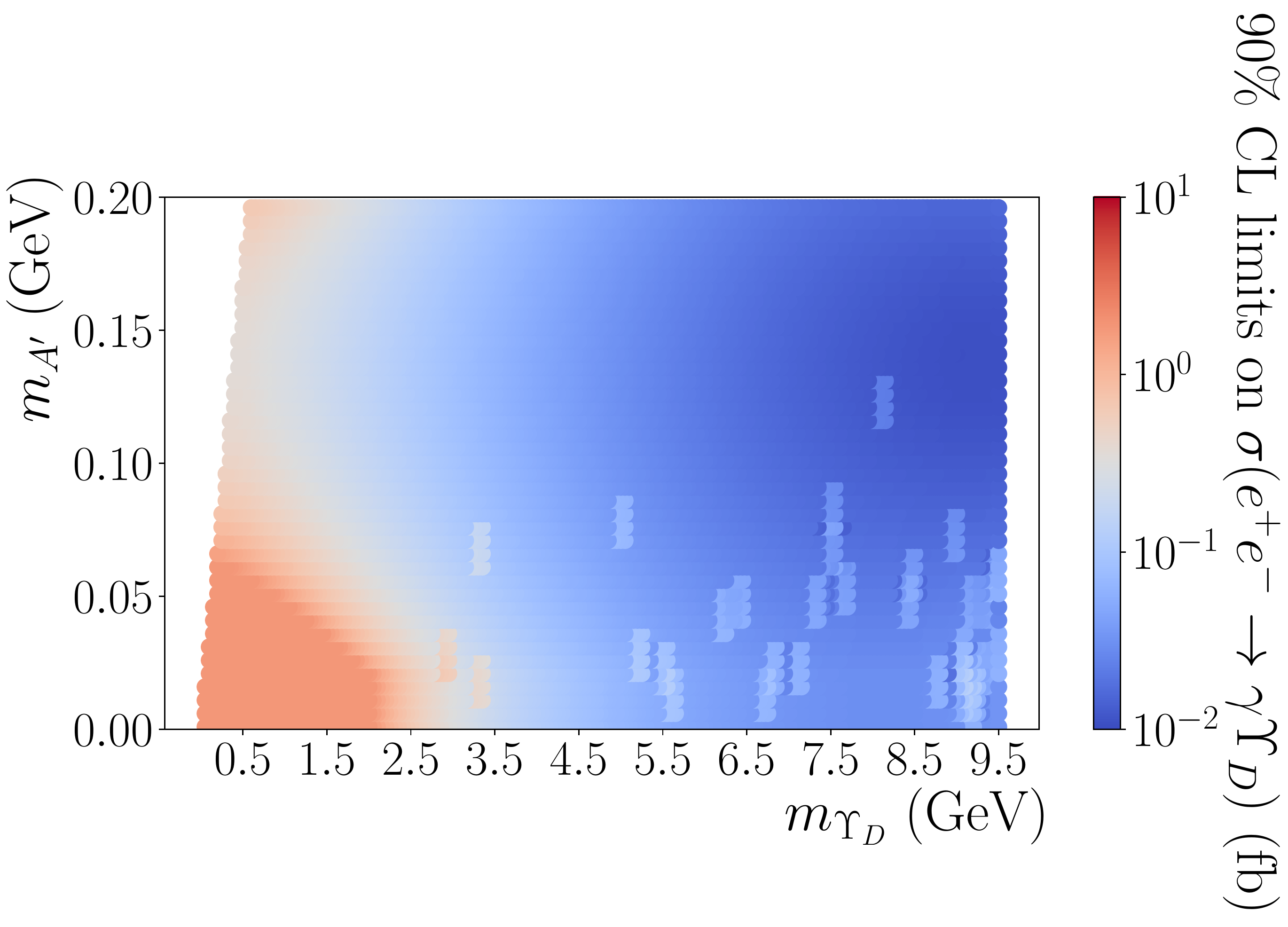}\\
\includegraphics[width=0.48\textwidth]{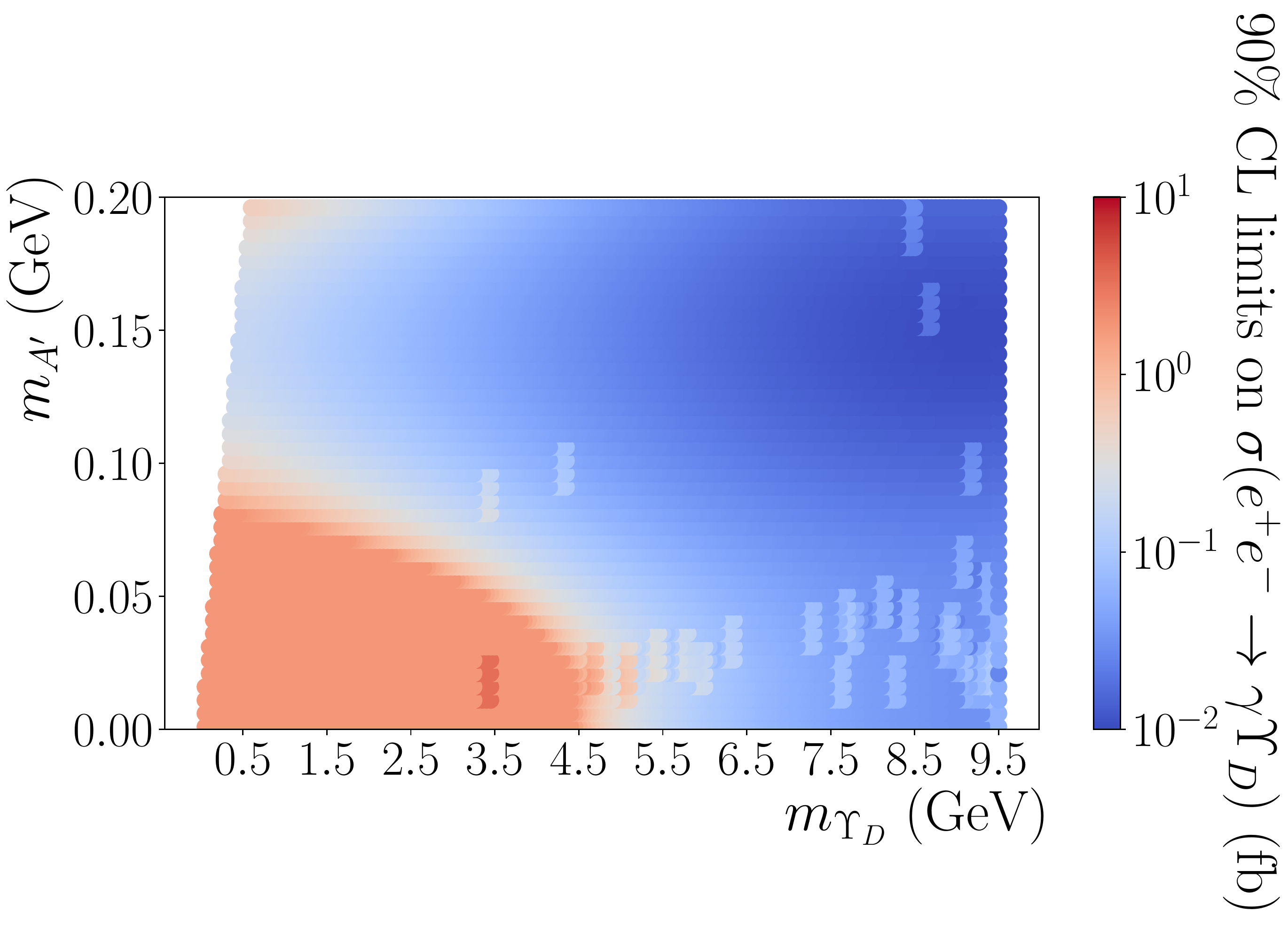}
\caption{The 90\% CL upper limits on the $e^+e^- \rightarrow \gamma \Upsilon_D$ cross section for dark photon 
lifetimes corresponding to (top left) $c\tau_{A'}=0.1 \rm \, mm$, (top right) $c\tau_{A'}=1 \rm \, mm$, and (bottom) 
$c\tau_{A'}=10 \rm \, mm$.}
\label{fig:cross_disp}
\end{figure}

\begin{figure}[htbp]
\centering
\includegraphics[width=0.48\textwidth]{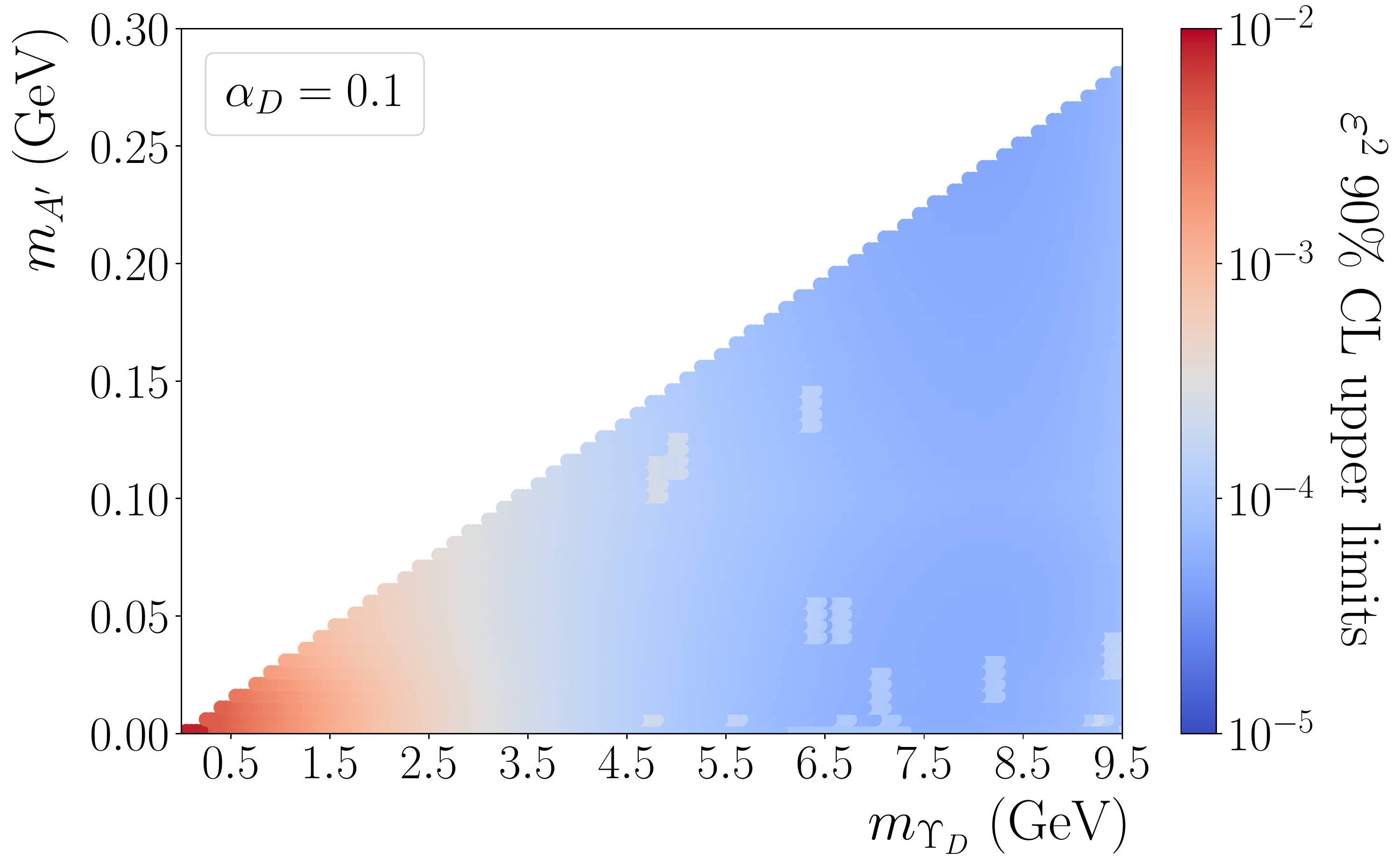}
\includegraphics[width=0.48\textwidth]{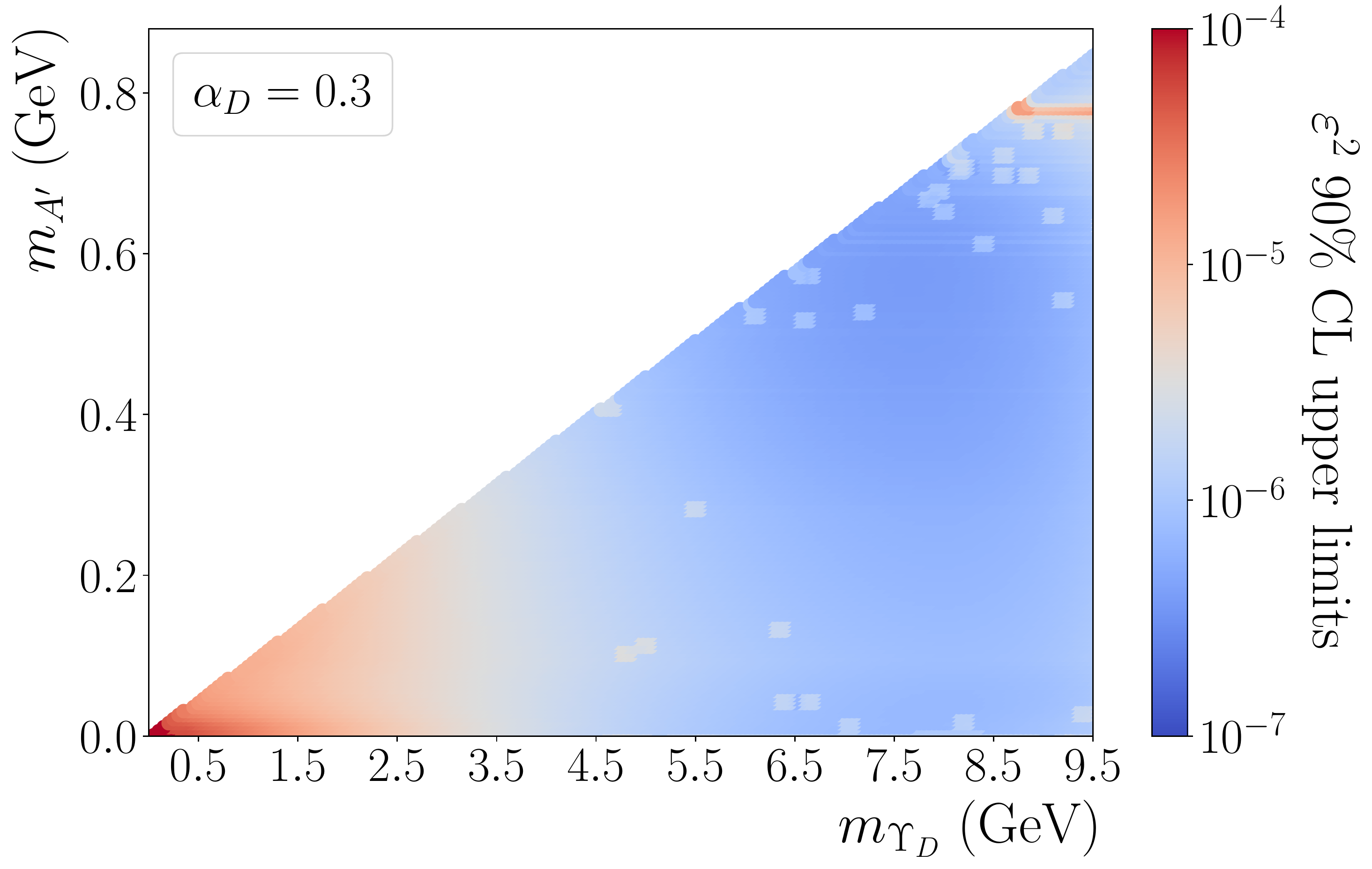}\\
\includegraphics[width=0.48\textwidth]{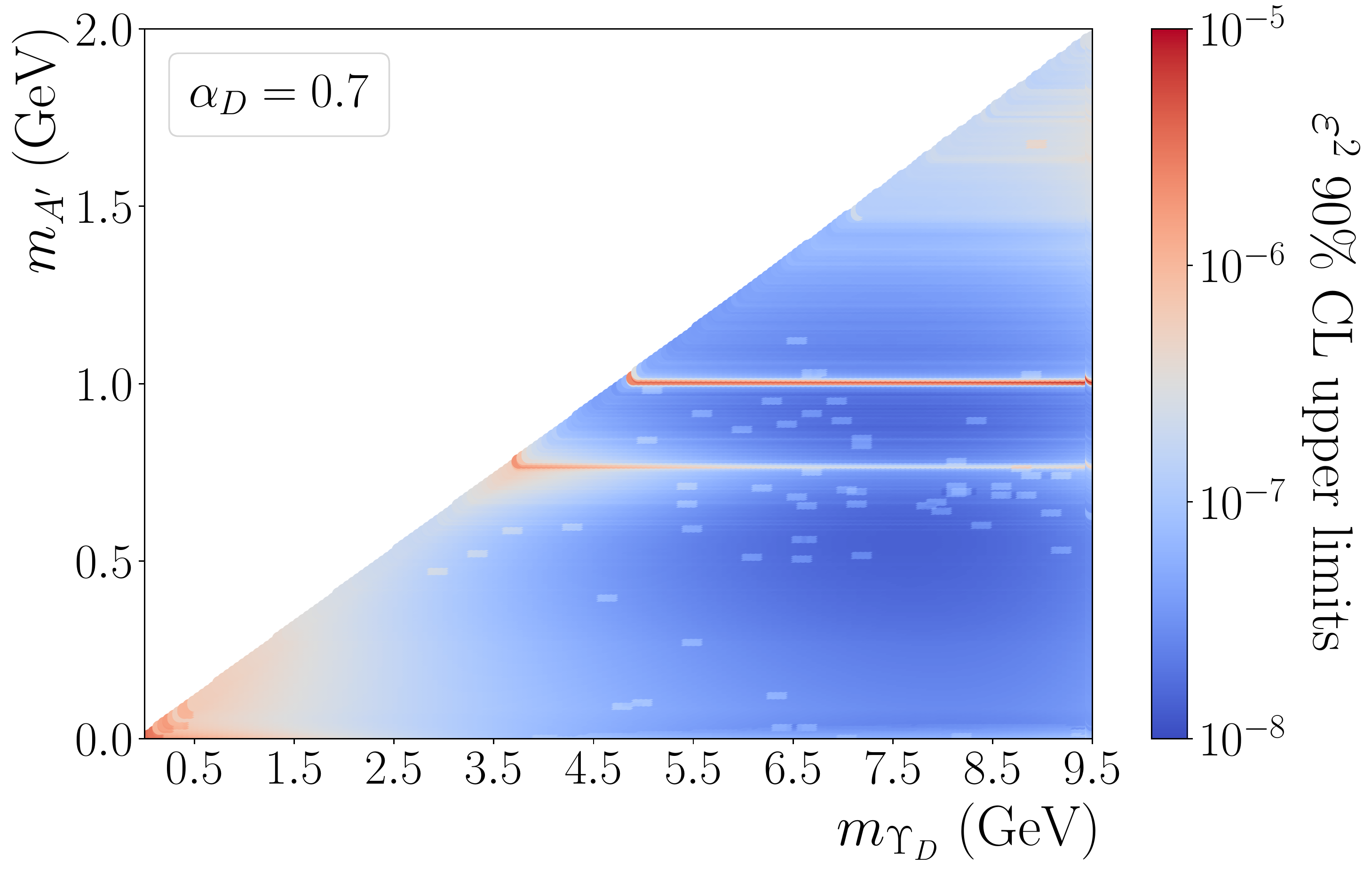}
\includegraphics[width=0.48\textwidth]{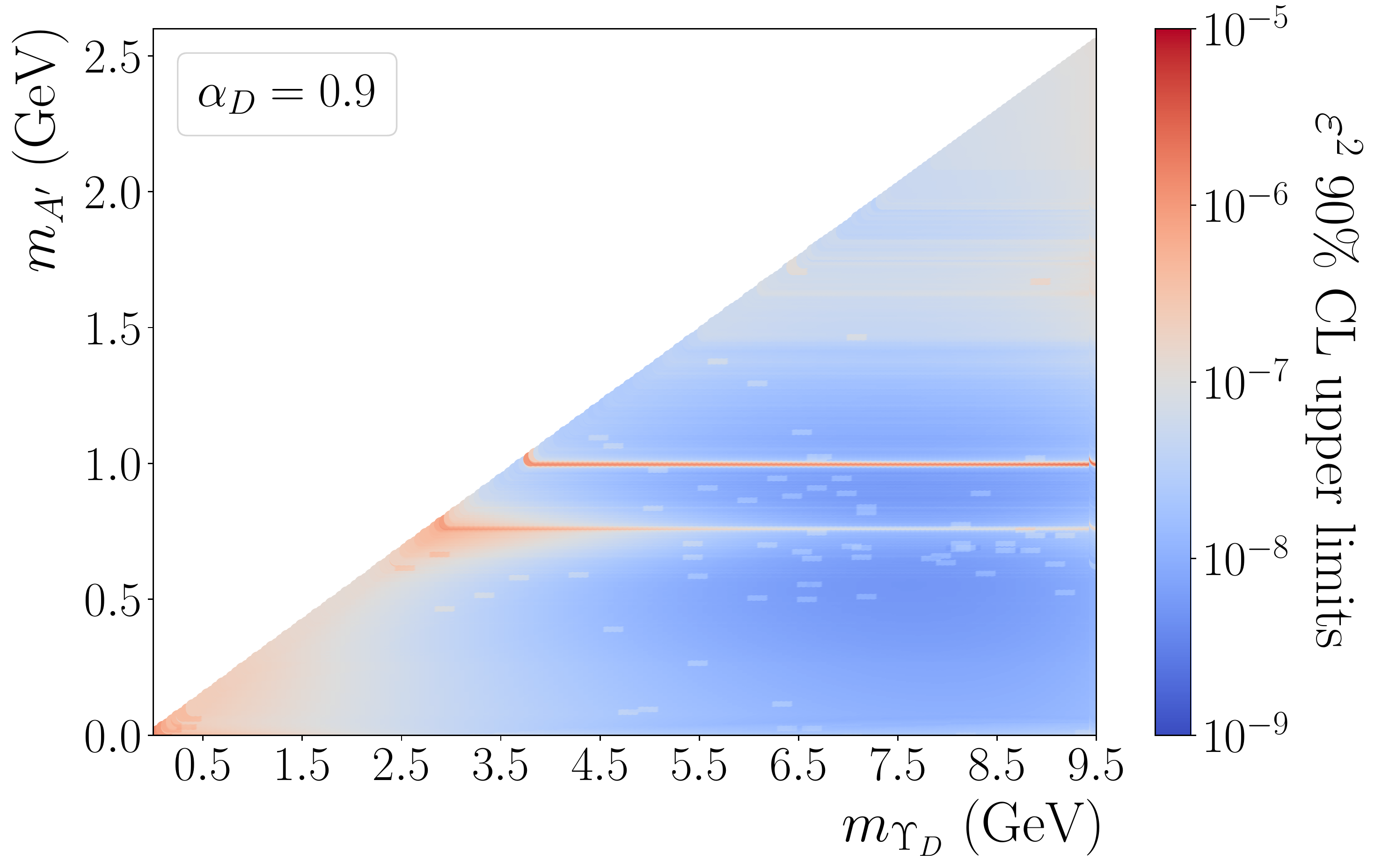}
\includegraphics[width=0.48\textwidth]{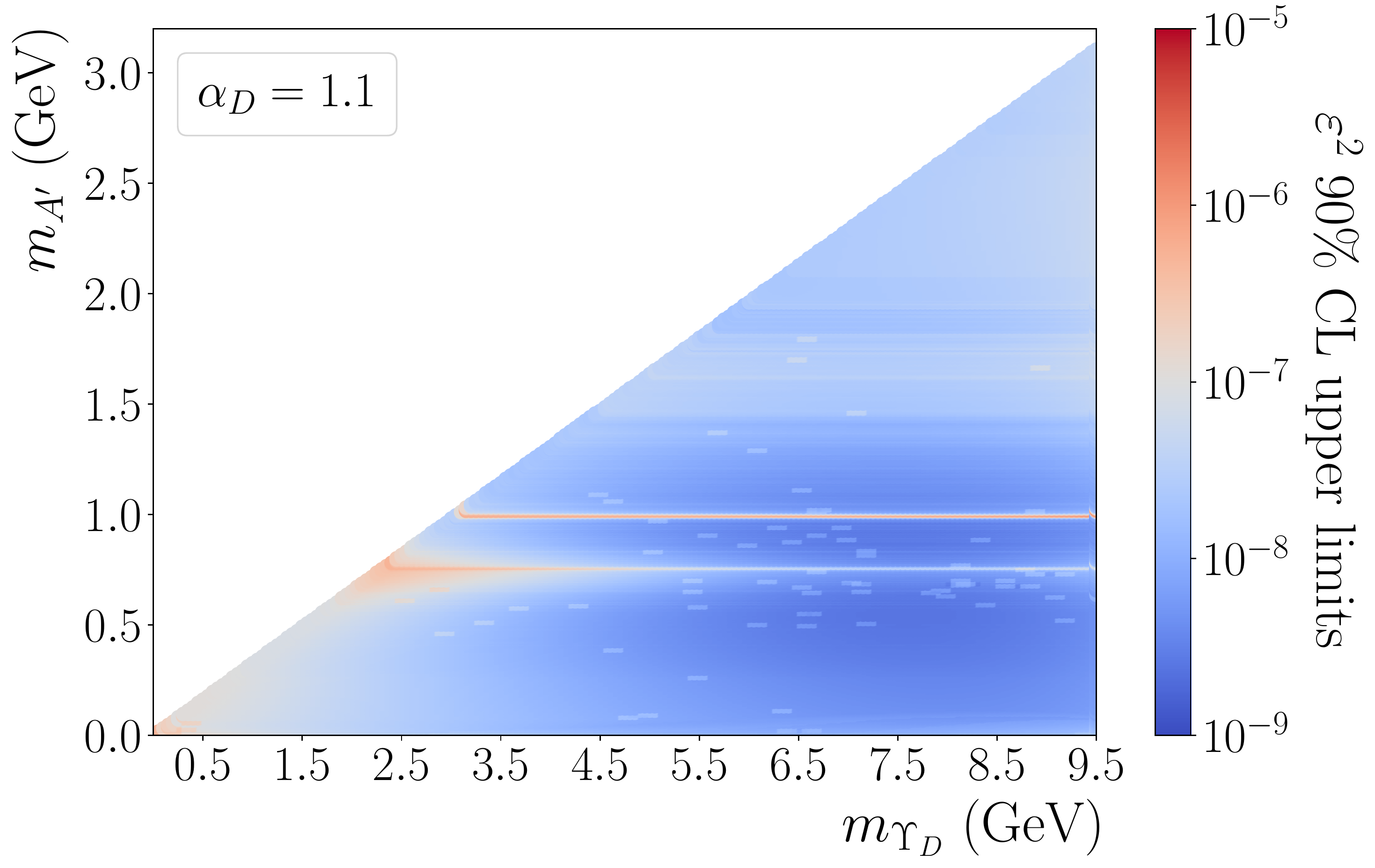}
\caption{The 90\% CL upper limits on the kinetic mixing $\varepsilon^2$ as a function of the $\Upsilon_D$ mass, $m_{\Upsilon_D}$, and 
dark photon mass, $m_{A'}$, assuming (top left) $\alpha_D = 0.1$, (top right) $\alpha_D = 0.3$, (middle left) $\alpha_D = 0.7$, (middle right) $\alpha_D = 0.9$, and (bottom) $\alpha_D = 1.1$.}
\label{fig:results3}
\end{figure}

\end{document}

%% file: authors_mar2021_frozen.tex
\author{J.~P.~Lees}
\author{V.~Poireau}
\author{V.~Tisserand}
\affiliation{Laboratoire d'Annecy-le-Vieux de Physique des Particules (LAPP), Universit\'e de Savoie, CNRS/IN2P3,  F-74941 Annecy-Le-Vieux, France}
\author{E.~Grauges}
\affiliation{Universitat de Barcelona, Facultat de Fisica, Departament ECM, E-08028 Barcelona, Spain }
\author{A.~Palano}
\affiliation{INFN Sezione di Bari, I-70126 Bari, Italy}
\author{G.~Eigen}
\affiliation{University of Bergen, Institute of Physics, N-5007 Bergen, Norway }
\author{D.~N.~Brown}
\author{Yu.~G.~Kolomensky}
\affiliation{Lawrence Berkeley National Laboratory and University of California, Berkeley, California 94720, USA }
\author{M.~Fritsch}
\author{H.~Koch}
\author{T.~Schroeder}
\affiliation{Ruhr Universit\"at Bochum, Institut f\"ur Experimentalphysik 1, D-44780 Bochum, Germany }
% ...
\collaboration{The \babar\ Collaboration}
\noaffiliation

%% file: acknowledgements.tex
We are grateful for the 
extraordinary contributions of our \pep2\ colleagues in
achieving the excellent luminosity and machine conditions
that have made this work possible.
The success of this project also relies critically on the 
expertise and dedication of the computing organizations that 
support \babar.
The collaborating institutions wish to thank 
SLAC for its support and the kind hospitality extended to them. 
This work is supported by the
US Department of Energy
and National Science Foundation, the
Natural Sciences and Engineering Research Council (Canada),
the Commissariat \`a l'Energie Atomique and
Institut National de Physique Nucl\'eaire et de Physique des Particules
(France), the
Bundesministerium f\"ur Bildung und Forschung and
Deutsche Forschungsgemeinschaft
(Germany), the
Istituto Nazionale di Fisica Nucleare (Italy),
the Foundation for Fundamental Research on Matter (The Netherlands),
the Research Council of Norway, the
Ministry of Education and Science of the Russian Federation, 
Ministerio de Econom\'{\i}a y Competitividad (Spain), the
Science and Technology Facilities Council (United Kingdom),
and the Binational Science Foundation (U.S.-Israel).
Individuals have received support from 
the Marie-Curie IEF program (European Union) and the A. P. Sloan Foundation (USA). 


%% file: paper.bbl
\begin{thebibliography}{9}
\bibitem{naturalUnits} Natural units ($\hbar=c=1$) are used throughout this paper.
\bibitem{helicity} The helicity angle is defined as the angle between the $\Upsilon_D$ momentum in the center-of-mass frame and the dark photon momentum in the $\Upsilon_D$ rest frame.
\end{thebibliography}
